\documentclass[12pt]{iopart}
\usepackage{array}
\usepackage{graphicx}

\def\be{\begin{equation}}
\def\ee{\end{equation}}
\def\bea{\begin{eqnarray}}
\def\eea{\end{eqnarray}}
\def\ba{\begin{array}}
\def\ea{\end{array}}
\def\bdm{\begin{displaymath}}
\def\edm{\end{displaymath}}

\begin{document}

\title[Current response of a topological insulator to a static Zeeman field]
{Current response of a topological insulator to a static Zeeman field}

\author{Bor-Luen Huang$^1$ and S.-K. Yip$^1$}

\address{$^1$ Institute of Physics, Academia Sinica, Taipei, Taiwan}

%\affiliation{Institute of Physics, Academia Sinica, Taipei, Taiwan}

%\date{\today }

\begin{abstract}
We study the magnetoelectric coupling at the surface of a topological insulator.
We are in particular interested in the surface current induced by a static Zeeman/exchange field.
This surface current can be related to the orbital magnetization of the system.
For an insulator with zero Chern number, the orbital magnetization is independent of the details at the boundary.
With the appearance of surface states in the topological insulator,
it is not immediately obvious if the response is not affected by the conditions at the surface.
We investigate this question using exact diagonalization to a lattice model.
By applying a time-reversal symmetry-breaking term near the boundary,
no matter if the surface states are gapped out, we still find no change in the surface current.
This arises from cancelations between Pauli and Van-Vleck contributions between surface and bulk scattering states.
We also show that the surface current response is independent of the chemical potential when it is within the bulk gap.
Our results are consistent with the claim that orbital magnetization is a bulk property.

\end{abstract}



\submitto{\JPCM}
%\submitto{\NJP}

\maketitle

\section{Introduction}

The discovery of topological insulator \cite{Hasan10,Qi11,Ando13} has generated
strong activities in the condensed matter and high-energy physics communities.
A prominent property of a topological insulator is the existence of gapless surface states
in the form of a Dirac Hamiltonian with momentum and spin directions locked with each other
\cite{Liu10,Zhang12}.
There have been many interesting theoretical predictions, e.g.
\cite{Raghu10,Nomura10,Burkov10,Garate10,Yokoyama14},
based on the current-spin coupling of this surface Hamiltonian.
If the Fermi energy is within the bulk gap, it is easy to expect that
some physical quantities or phenomena, such as
the Landau-level spectrum in a perpendicular field \cite{Liu10,Cheng10,Hanaguri10},
transport phenomena involving the surface states \cite{Lu14}, or
the RKKY interactions \cite{Liu09,Zyuzin14} between spins on the surface,
can be evaluated from an effective surface Hamiltonian.
On the other hand, one may question if some physical quantities
do not just rely on effective surface models \cite{Black-Schaffer12}.
The response induced by an external magnetic field is one of the interesting topics with above ambiguity.
Generally, this perturbation to a system has an effective Zeeman field term,
a vector potential term, and further contributions \cite{Roth62,Wannier64,Buot72}.
In this paper, we focus on the surface current response to a static Zeeman or exchange field.
We examine if the surface Hamiltonian is sufficient to determine this response.
We also can relate the current ${\bf j}$ to the orbital magnetization ${\bf M}$
by ${\bf M}=\frac{1}{2}\int {\bf r}\times{\bf j}({\bf r})d{\bf r}$ \cite{Jackson98},
where ${\bf r}$ is the position.

Orbital magnetization is an interesting quantity.
Even when the maximally localized Wannier functions are used, this magnetization,
besides a term (``local circulation" in \cite{Thonhauser05,Ceresoli06} or ``self-rotation" in \cite{Souza08})
which can be interpreted in the same way as rotational motion of electrons in isolated atoms,
contains a contribution (``itinerant circulation" in \cite{Thonhauser05,Ceresoli06,Souza08})
which describes the electronic motion at the edge of the sample.
Despite the existence of such a current near the sample boundary,
it has been shown that the total orbital magnetization is independent of the details at
the boundary for an insulator with zero Chern number \cite{Thonhauser05,Ceresoli06}.
With the presence of topological surface states for Chern insulators,
this independence becomes even less obvious.
Nevertheless, Bianco and Resta \cite{Bianco13} provided a real-space expression of
orbital magnetization for any two-dimensional insulator with finite Chern number, and
demonstrated that the formula is independent of boundary conditions for a large sample.
On the other hand, Chen and Lee \cite{Chen12} argued that,
for a system composed of two insulators with opposite Chern numbers, the orbital
magnetization should be independent of the details at the boundaries.
They also demonstrated, by numerical calculations,
that the orbital magnetization is insensitive to local perturbed potentials
near the edge in a Chern insulator, and it is also unchanged
with magnetic moments at the edges, which gap out the edge states, in spin Hall insulator.
Therefore, it is interesting to verify if this bulk property still applies
for a three-dimensional topological insulator, where there are topologically
required surface states with the low energy physics described by two-dimensional Dirac Fermions.

In an earlier work, one of us \cite{Yip14} analyzed the surface current of a topological insulator generated
by an in-plane static Zeeman field via perturbation theory, employing an effective
Hamiltonian that is valid only for small bulk momentum $\vec k$.
He showed that there is a cancelation between the contribution from redistribution of particles (Pauli) and
that from virtual transitions (Van-Vleck) between the surface states.
In particular, due to this cancelation, the current response is independent of the chemical potential
when it falls within the bulk gap.
Furthermore, it is found that the bulk scattering states also provides a significant contribution
of opposite sign to that of the surface states.
As a result of this cancelation, the total surface current has no dramatic change when
the system changes from a trivial to a topological insulator.

In this paper, to account more carefully contributions from states of all momenta,
we calculate the current response to a static external Zeeman field, employing a lattice model.
Using exact diagonalization, we examine the contributions from both the surface and the scattering states.
To gain more physical insights, we shall consider in some detail the contributions from states with
small momenta parallel to the surface and compare these results with those obtained analytically in \cite{Yip14}.
The cancelation between the Pauli and the Van-Vleck contributions for the surface
states, and the Van-Vleck contributions from the surface versus the bulk states, will be examined more carefully.
We compare the current responses between topological insulators and normal insulators.
For the topological insulator, we demonstrate the independence of the response on the chemical potential
(so long as it is within the bulk gap), even though the occupation of the surface states depends on it.
We shall also show that, due to compensation between the surface and
scattering states contributions, the total current response is independent of
surface magnetic moments which gaps out the surface states.
We shall also discuss effects of different types of external fields on different surfaces.
We consider anisotropic topological insulators and examine
the origin of the anisotropy in the current response.
Some technical difficulties we encounter in using the exact diagonalization to evaluate
the Pauli contributions and the contributions near the Dirac point are mentioned.

This paper is outlined as follows. In Sec. \ref{model}, we describe the theoretical model
being used. The numerical results for the current in
response to an in-plane external field are provided in Sec. \ref{current-y}.
We compare the results obtained here with the analytic study of \cite{Yip14}.
We also study the effects of chemical potential, anisotropy, and surface magnetic moments.
In Sec. \ref{field-type}, cases with different types of external fields acting on different surfaces
are discussed. Sec. \ref{conclusion} is the conclusion.

\section{Model Hamiltonian}\label{model}

An effective model for a topological insulator, for example Bi$_2$Se$_3$,
at small momentum $\vec k$ is of the form \cite{Liu10}
\be
H=m\sigma_{x}+v_{z}k_{z}\sigma_{y}-v(k_{x}s_{y}-k_{y}s_{x})\sigma_{z}.
\label{hb-linear}
\ee
Here $s_{i}$ and $\sigma_{i}$ are the Pauli matrices acting on spin and
orbital subspaces, respectively. $m$, $v$ and $v_{z}$ are material parameters.
$v$ and $v_{z}$ are set to be positive. The parity operator is taken as $\sigma_{x}$.
To include the contribution from finite momenta, we extend the Hamiltonian Eq. (\ref{hb-linear})
to the following form:
\be
H=m'(k)\sigma_{x}+v_{z}\sin(k_{z})\sigma_{y}-v(\sin(k_{x})s_{y}-\sin(k_{y})s_{x})\sigma_{z},
\label{hb}
\ee
where $m'(k) = m + c (2-\cos(k_x)-\cos(k_y)) + c_z (1-\cos(k_z))$.
For convenience, we set the lattice constant ($a$) equal to one. $c$ and $c_{z}$
are also material parameters and limited to be positive. $m > 0$ specifies a trivial phase.
The condition to be in the topological insulator depends on both $m$, $c$ and $c_z$.
We shall focus on the topological non-trivial phase within the following region,
\be
-2\min\{c,c_{z}\}<m<0.
\label{hb-sscondition}
\ee
Here for simplicity,
we have written down a form that is appropriate for a crystal of tetragonal $D_{4h}$
symmetry instead of the $D_{3d}^{5}$ symmetry for the real Bi$_{2}$Se$_{3}$ family.
We make this simplication because the relevant issues of the current response in this paper
is whether the contribution is from the surface states or from the scattering states,
hence our main conclusions are independent of the lattice symmetry and
the specific parameters of the materials.

We shall start from lattice model to take account of effects of boundaries.
The corresponding real-space Hamiltonian for a simple cubic lattice is
\be
\begin{array}{lll}
H(r) & = & (m+2c+c_z)\sum_{r}a^{+}_{r}\sigma_{x}a_{r} \\
  & + & \sum_{r}[ a^{+}_{r}(-\frac{c}{2}\sigma_{x}+i\frac{v}{2}s_{y}\sigma_{z})a_{r+\delta x} \\
  &   & +a^{+}_{r}(-\frac{c}{2}\sigma_{x}-i\frac{v}{2}s_{x}\sigma_{z})a_{r+\delta y} \\
  &   & +a^{+}_{r}(-\frac{c_{z}}{2}\sigma_{x}-i\frac{v_{z}}{2}\sigma_{y})a_{r+\delta z} + h.c. ],
\end{array}
\label{hb-lattice}
\ee
where $a^{+}_{r}$ and $a_{r}$ are Fermion creation and annihilation operator, respectively,
at site r. To calculate the surface current, for example, on the top surface with the normal
along $+z$ direction, we shall apply an open boundary condition to the $z$ direction
and periodic boundary conditions to the $x$ and $y$ directions.
In this case, we get an effective Hamiltonian on an one-dimensional chain along
the $z$ direction with two good quantum numbers, the in-plane momentum components $k_{x}$ and $k_{y}$.
We get the energy spectra from the exact diagonalization.
For the topological trivial phase, the spectra consist simply of scattering states,
related to the bulk plane-wave states found in a system without boundaries and occupying the same
energy range.  For the topological non-trivial phase, there are also mid-gap surface states.
These midgap states form a Dirac cone for the top ($+z$)
surface and another for the bottom ($-z$) surface.
The energy dispersion of the surface states is given by
$E_{s}=\pm v\sqrt{\sin^{2}(k_{x})+\sin^{2}(k_{y})}$, which can be related
to zero energy states of its corresponding supersymmetric Hamiltonian. \cite{Huang04}
The condition to have surface states is $|m_{k}^{z}|<|c_{z}|$, where $m_{k}^{z} = m'(k_{z}=\pi/2)$.
Combining the condition $|m_{k}^{x}|<|c|$ for the side surface in $+x$ direction, where
$m_{k}^{x} = m'(k_{x}=\pi/2)$, we shall set $m$ within Eq. (\ref{hb-sscondition})
for simplification.
Note that low energy dispersion at half-filling is described by massless Dirac Fermions.
For a given in-plane momentum, the decay length of the surface state is
$\lambda=|\ln|m_{k}^{z}/c_{z}||^{-1}$ in the z direction, which reduced to
$v_{z}/|m|$ as $c_{z}=v_{z}$ and $|m|\ll v_{z}$.
In this paper, we shall focus on the case with the thickness much larger than $\lambda$
so that the coupling between the surface states on different surfaces can be ignored.

\section{Current response to an in-plane Zeeman field}\label{current-y}

When applying an in-plane Zeeman field $B_{y}$ to a system with spin-orbit coupling, one can
expect redistribution of particles around the Fermi surface because these states
have spin-momentum locking and the energy levels become lower or higher
as the spin prefers to be along or against the external field. Therefore there
is a contribution to the current in the transversal direction of the field. This is the
Pauli contribution of the current response. However, the Zeeman field also
modifies the wavefunctions. The modification will trigger off virtual transitions
between occupied and empty states \cite{Yip02}.
This is the Van-Vleck contribution of the current response.
As mentioned in Ref. \cite{Yip14}, the virtual transition between the surface states
is not enough to get a physical answer for the current response to an external field,
since the result depends on the momentum cut-off employed.
A proper evaluation of the Van-Vleck contribution must consider virtual transitions
from any occupied state to any empty state.
Here we use exact diagonalization to evaluate all the states and
the current they carry. The operator for current density along the $x$ direction is given by
\be
J_{x}(z,\vec{k}_{\parallel})=\partial H(z,\vec{k}_{\parallel})/\partial k_{x},
\ee
where $H(z,\vec{k}_{\parallel})$ is the Hamiltonian after partial Fourier transformation as
described below Eq. (\ref{hb-lattice}). Therefore, for the model Eq. (\ref{hb}),
this is given by $c \sin(k_{x})\sigma_{x}-v \cos(k_{x})s_{y}\sigma_{z}$.
We obtain the eigenstates with finite external field $B_{y}$ by exact diagonalization
as mentioned before, and evaluate the expectation value of this current operator.
For the external perturbation to create the surface current, we use $H_{B1,y}=-g_{1}s_{y}B_{y}$
where $g_{1}$ is also a material parameter and would be absorbed in the definition for $B_y$ for simplicity.
The most general perturbation by an external Zeeman field $B_{y}$ has two types \cite{Liu10},
with and without an additional $\sigma_{x}$ coefficient in the above expression.
We shall see that the term with $\sigma_{x}$ contributes no current response to $B_{y}$ on the top surface.
However, it becomes necessary for other cases.
We have more discussion about this point in Sec. \ref{field-type}.
We shall consider systems with a uniform $B_{y}$ at first, and discuss the effects
of non-uniform $B_{y}$ in Sec. \ref{non-u}.
For a uniform $B_y$, the current deep in the bulk vanishes, since Eq. (\ref{hb-lattice})
for the bulk has inversion symmetry.  The current flows only near the surface.
The surface (number) current on the top surface can then be evaluated by
\be
I_{top}^{x}=\int_{B.Z.} \frac{ d^{2}\vec{k}_{\parallel}}{4 \pi^2}
\sum_{z\in \mbox{top half}}J_{x}(z,\vec{k}_{\parallel}),
\label{Itop}
\ee
where $\vec k_\|$ is the momentum parallel to the surface.
The surface current for the bottom surface is from similar formula but the sum
$z$ is over the bottom half.
Although we could have studied general responses, we shall confine ourselves to
the linear case where physical pictures can be accessed more easily.  We therefore
present our results in the form of a
linear response coefficient defined by
\be
\kappa=I_{top}^{x}/ B_{y}.
\label{kappa}
\ee

\subsection{Surface and scattering states contributions}\label{p-v}

\subsubsection{Chemical potential:}

First of all, we shall show that the current response to $H_{B1,y}$ is independent of the chemical potential,
provided that it is within the bulk gap.
In this subsection, the chemical potential independence is verified from numerical results.
Combining with other studies below, we shall also give a physical argument
why this response is independent of the chemical potential before the end of Sec. \ref{smm}.

\begin{figure}
\begin{center}
\includegraphics*[width=150mm]{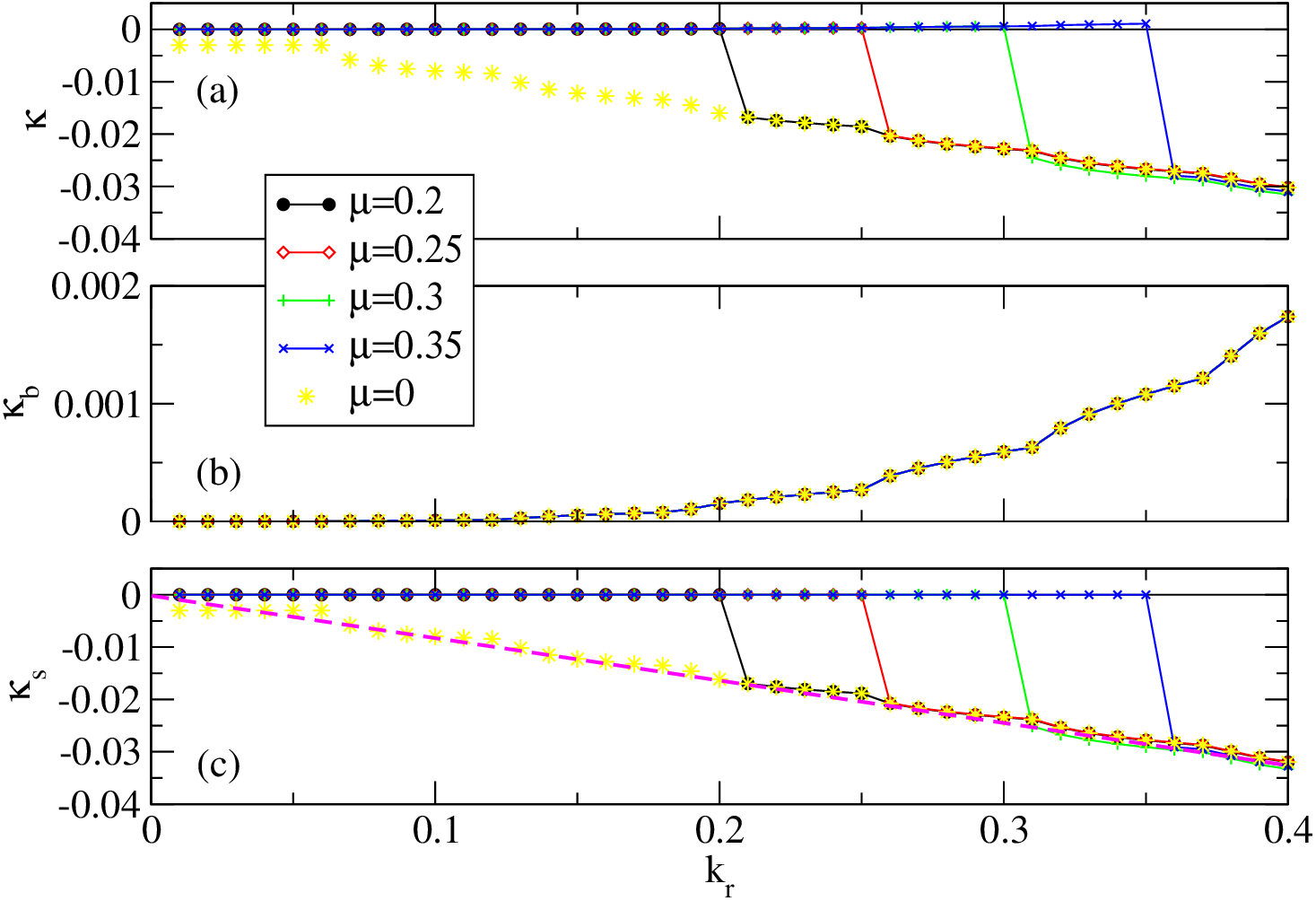}
\caption{(Color online) Current response of a topological insulator from states with
$k_\| < k_{r}$. (a) Total contributions.
(b) Contributions from the bulk scattering states.
(c) Contributions from the surface states.
$c=v=c_{z}=v_{z}=1$.
$\kappa_{b}$'s are the same for different $\mu$'s.
A sharp change of $\kappa_{s}$ around the Fermi energy is corresponding to the Pauli contribution.
We limit the chemical potential to be $0\leq\mu<|m|$, where
$m=-0.4$ in these plots.
The magenta dash line is linear fitting around small k for the case with $\mu=0$.
}\label{kappa-kr}
\end{center}
\end{figure}

In order to make the physics more transparent, we shall first compare the results from numerical calculations
with those from analytic works \cite{Yip14} by studying the response of the states with small $\vec k_\|$.
We do this by limiting the %integral
domain in Eq. (\ref{Itop}) to $k_\| < k_r$:
\be
\int_{B.Z.} d^{2}\vec{k}_{\parallel} \rightarrow 2\pi\int_{0}^{k_r}dk_{\parallel},
\ee
where $k_\| \equiv | \vec k_\| |$ is
the magnitude of the in-plane momenta.
We divide the total current response ($\kappa$) into two contributions, one from the surface states
(denoted by $\kappa_{s}$) and one from the bulk bands (denoted by $\kappa_{b}$).
The results are shown in Fig. \ref{kappa-kr}.
We limit the chemical potential to be $|\mu|<|m|$, so that the system
has no scattering states around the Fermi energy.
Figure \ref{kappa-kr}(b) shows that $\kappa_{b}$ is independent of $\mu$, though it is finite.
This independence is in accordance with \cite{Yip14}, where it was concluded that
there is no contribution to the current via virtual transitions from
the valence band to the surface states and from the surface states to the conduction band.
That is, $\kappa_b$ arises entirely from virtual transitions between the valence and conduction bands.
$\kappa_{s}$ are shown in Fig. \ref{kappa-kr}(c).
For $\mu = 0$, there is no Pauli contribution due to the vanishing of density of states, and $\kappa_s$ is
entirely due to virtual transitions from occupied to empty surface states.
The magnitude of $\kappa_s$ increases with $k_r$ when more surface states are included, as shown by the dashed line
(Our actual numerical calculation gives the yellow dots with some small deviations from this line
near small $k_r$.  This is related to a technical issue which will be discussed in Sec. \ref{Tech}).
For finite $\mu$'s, $\kappa_{s}$ becomes zero when $k_{r}$ is less than the Fermi momentum $k_{F}$.
This reflects the absence of virtual transitions between the surface states
when $k_r < k_F$ due to the Pauli exclusion principle, since at a given momentum $\vec k_\|$ with $k_\| < k_F$,
both the surface states with $E_s {> \atop <} 0$ are occupied.
Around $k_r \sim k_{F}$, the response has a sharp change, which is mainly from redistribution
of particles by the external field. This represents the Pauli contribution.
The finite width of the change in $\kappa_s$ in Fig. \ref{kappa-kr} is due both to the fact that
the Fermi surface is not a regular circle and the finite the resolution of the points that we have used.
This Pauli contribution complements exactly the missing Van-Vleck contributions
(compared to $\mu = 0$) from surface states with $k_\| \leq k_F$.
$\kappa_s$ becomes independent of $\mu$ when $k_r > k_F$.
$\kappa = \kappa_s + \kappa_b$ is given in Fig. \ref{kappa-kr}(a).
At these small momenta, $\kappa_{s}$ is main contribution to $\kappa$, while $\kappa_{b}$
is a smaller contribution of opposite sign.
However, as shown in Fig. \ref{kappa-aniso}, $\kappa_{b}$ shows non-monotonic behavior
and has comparable amplitude to that of $\kappa_{s}$ for some cases.
The comparison between Fig. \ref{kappa-kr} and Fig. \ref{kappa-aniso} for isotropic case
also shows that $\kappa_{b}$ versus $\kappa_{s}$ differ more (less) for increasing (decreasing) $|m|$.
The numerical results for small $k_{r}$ discussed above is consistent with the analytic results in \cite{Yip14}.
The current response is insensitive to the occupancy of the surface states, which
infers it is a quantity related to the bulk properties.

Numerical results also show that $\kappa$ for the bottom surface has equal amount but opposite sign
to the top surface.
This is related to the charge conservation.
We shall provide an argument in Sec. \ref{smm} why this current
must be independent of the chemical potential. \footnote{%\cite{footnote-C}
An alternate way to understand this result is as follows.
As mentioned in Sec. \ref{LM},
we can consider our system as a collection of two-dimensional systems, one for each
$k_y$ (for a magnetic field along $y$).  In our case, each of these two-dimensional systems
has zero Chern number (at finite $B_y$, as can most easily be seen by considering the surface state
spectra).   On the other hand, the derivative of the orbital magnetic moment with respect
to the chemical potential is proportional to the Chern number for a general two-dimensional
system \cite{Ceresoli06}.  Hence this derivative must also vanish for our three-dimensional
system.}

\subsubsection{Dependence on Hamiltonian parameters:}

\begin{figure}
\begin{center}
\includegraphics*[width=100mm]{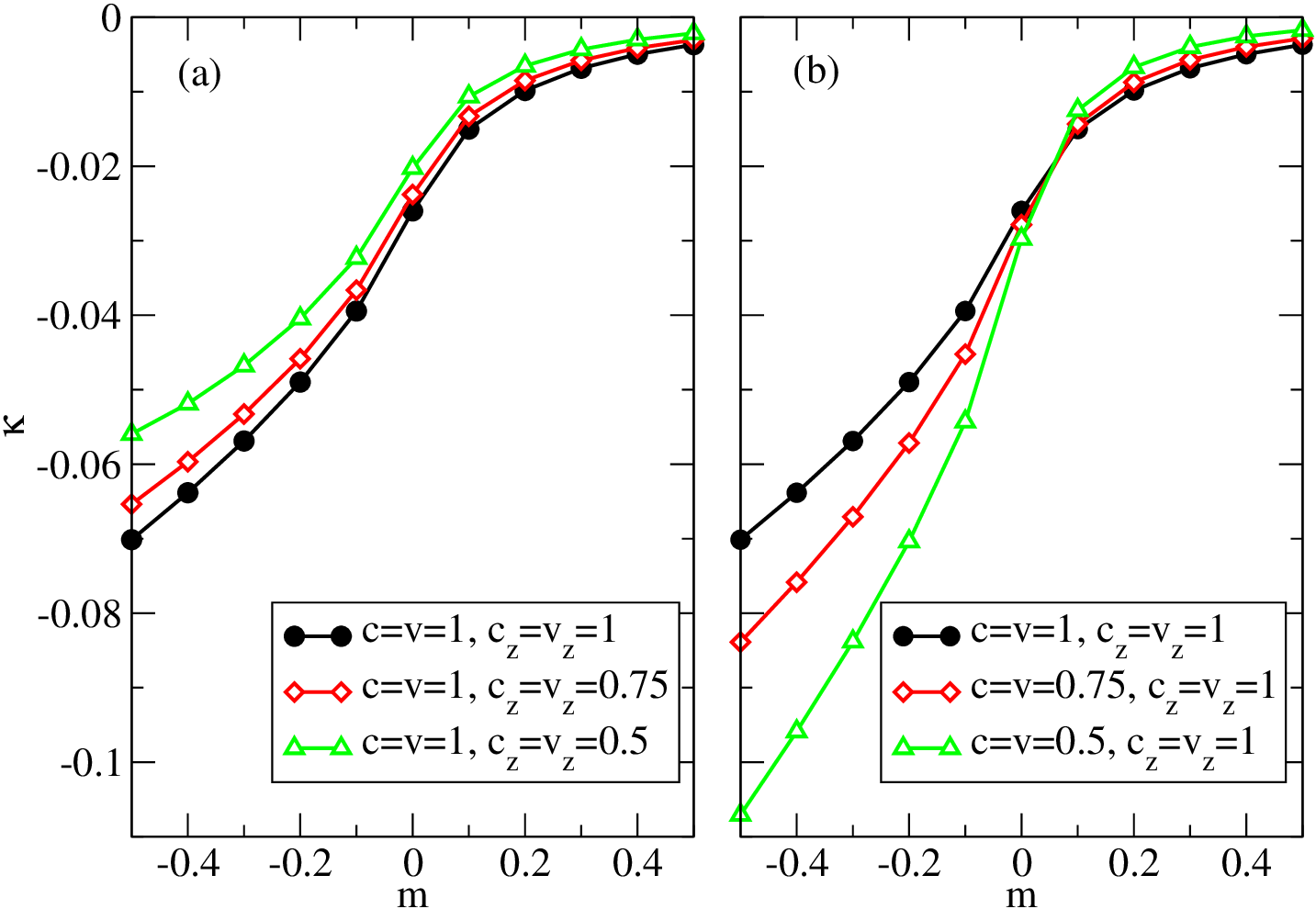}
\caption{(Color online) Current response for different $m$'s to $B_{y}$ on the top surface.
(a) Fixed $c$ and varying $c_{z}$. (b) Fixed $c_{z}$ and varying $c$.
Total current responses on the side surface are found to be the same as those on the top surface.
}\label{kappa-m}
\end{center}
\end{figure}

\begin{figure}
\begin{center}
\includegraphics*[width=100mm]{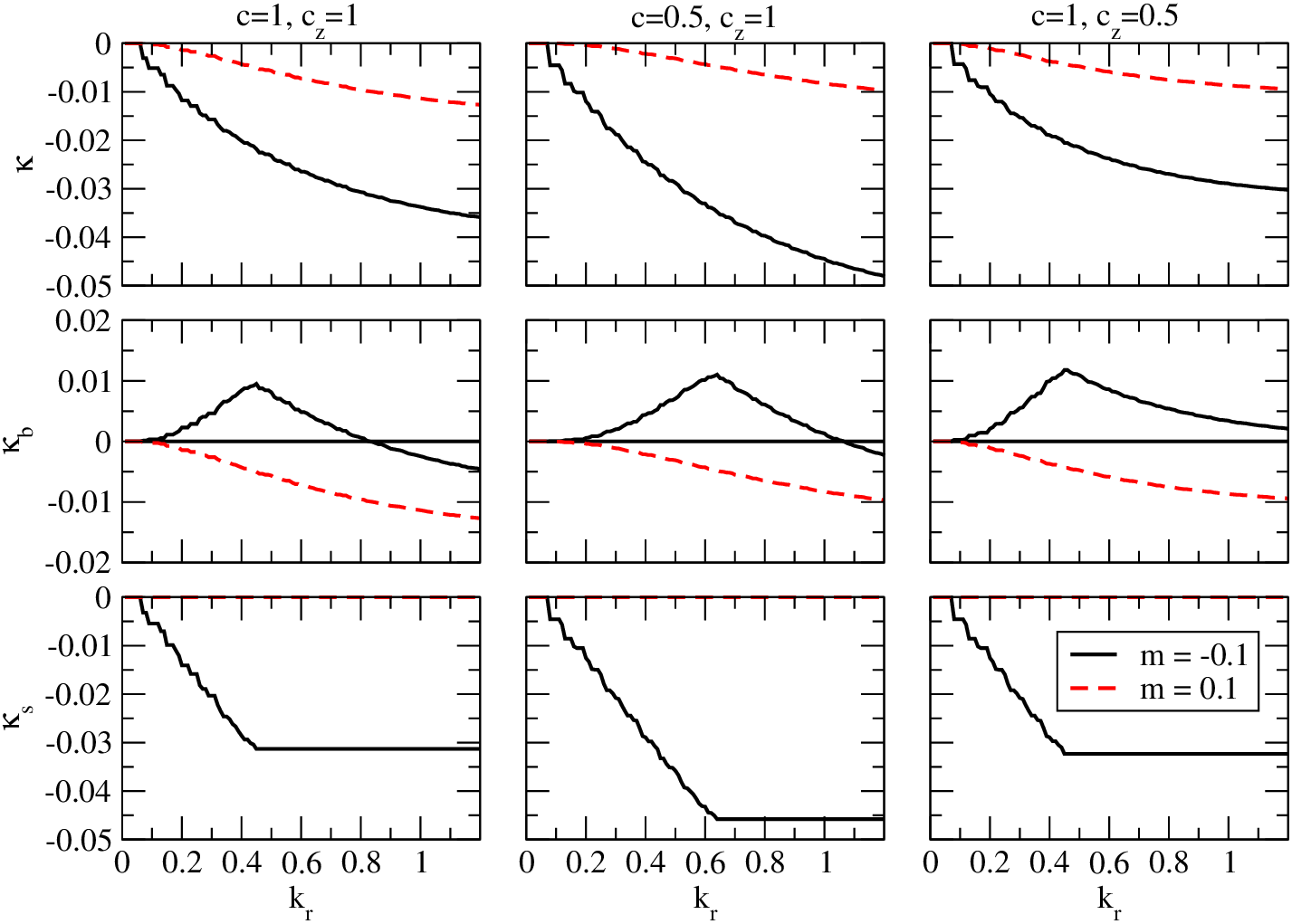}
\caption{(Color online) Anisotropy study for topological (black line) and trivial insulator
(broken red line) on the top surface. The external field is along the $y$ direction.
Three cases with different material parameters are shown in different
columns. Top row is for total current response, middle row for the contribution from
scattering states and the bottom for that from surface states. Black full lines
are for topological non-trivial phase and red dash lines for trivial phase.
}\label{kappa-aniso}
\end{center}
\end{figure}

The total surface current responses are dependent on the material parameters, i.e.
$m$, $c$, $c_{z}$, $v$ and $v_{z}$. We like to show more about this point.
For convenience, we set $c=v$ and $c_{z}=v_{z}$.
Because of the independence of $\kappa$ on the chemical potential,
we shall study the systems with $\mu=0$.
For a system with $c\neq c_{z}$, it means the anisotropy between in-plane and out-of-plane.
In Fig. \ref{kappa-m}, we show that the system has different current responses
and anisotropic effects between topological and normal insulators.
In the normal phase, non-zero current from only the scattering states
is slightly modified as changing the material parameters.
As $c$ or $c_{z}$ become larger, the responses are larger.
When $m$ changes from positive to negative values, the response
continually grows up into the topological phase. Instead of weak dependence and
monotonic increasing for the current response of normal insulator,
it shows different dependence on $c$ and $c_{z}$ in the topological phase.
Figure \ref{kappa-aniso} shows more details about the anisotropic effects.
Note that the external field is set to be so small that the surface states
are well separated from the bulk bands for all momenta.
In the normal phase, the current response is only from the scattering states
and enhanced for larger $c$ or $c_{z}$, as shown in the middle row of Fig. \ref{kappa-aniso}.
The amplitude of the response is monotonic increasing as a function of $k_{r}$ for small momenta.
For the topological phase, the dependence is more complicated.
First, we focus on $\kappa_{s}$ in the bottom row of Fig. \ref{kappa-aniso}.
Before saturation, the slopes of $\kappa_{s}$ are the same for all cases,
which related to the independence of $c$ for the Van-Vleck contribution. \cite{Yip14}
We also find that total $\kappa_{s}$ is weakly changed as modifying $c_{z}$, but strongly dependent on $c$.
This might be related to the fact that the size of surface cone becomes smaller as $c$ is larger.
However, the results of $\kappa$ (top row of Fig. \ref{kappa-aniso}) show that it is still dependent on $c_{z}$,
which mainly related to $\kappa_{b}$ (middle row of Fig. \ref{kappa-aniso}).
$\kappa_{b}$ shows an opposite contribution for those momenta with the surface states.
Once the surface states merge into bulk bands, $\kappa_{b}$ is changing
in a way similar to that of normal insulator.
We note that $\kappa$ increases smoothly as crossing the transition, which infers that
$\kappa_{b}$ compensates the missing of the midgap states at larger momenta.
Basically, the momentum region to have surface cones is a bulk property.
In addition, $\kappa$ is independent of the chemical potential as $|\mu|<|m|$.
Therefore, the current response to an external field cannot be described only from
effective surface Hamiltonian.

\subsubsection{Technical remarks:}\label{Tech}

Before further discussions, we like to mention some technical problems we encounter
when calculating the current by the exact diagonalization, and how they are overcome.
The first is the difficulty to calculate
the contribution with momenta near the Dirac point as $\mu=0$.
Due to finite size effect, two surfaces couple to each other and result in a gap,
inversely exponentially proportional to the size of the sample, to the ideal Dirac cones.
This causes a sudden jump for the current as the energy scale of the external field
is larger than this gap. The corresponding current response will reflect size effects
and this is already beyond the linear response, which is not the concern in this paper.
From a physical point of view, the density of state approaches zero at the Dirac point,
and so the Pauli contribution should vanish. In the numerical calculations,
we either subtract out the spurious contributions to the current
(when larger fields were used in the exact diagonalization),
or slightly shift the momentum to avoid the ambiguity.
Another problem is the difficulty to count the Pauli contribution for cases with $\mu\neq 0$.
Finite grid makes the evaluation of redistribution of particles difficult.
We can overcome this problem by applying twist boundary conditions and averaging over
different cases. This method is equivalent to including more lattice momentum points.
The twist boundary condition can also be applied to estimate the contribution
around the Dirac point.

\subsection{Relation to 2D system}\label{LM}

\begin{figure}
\begin{center}
\includegraphics*[width=100mm]{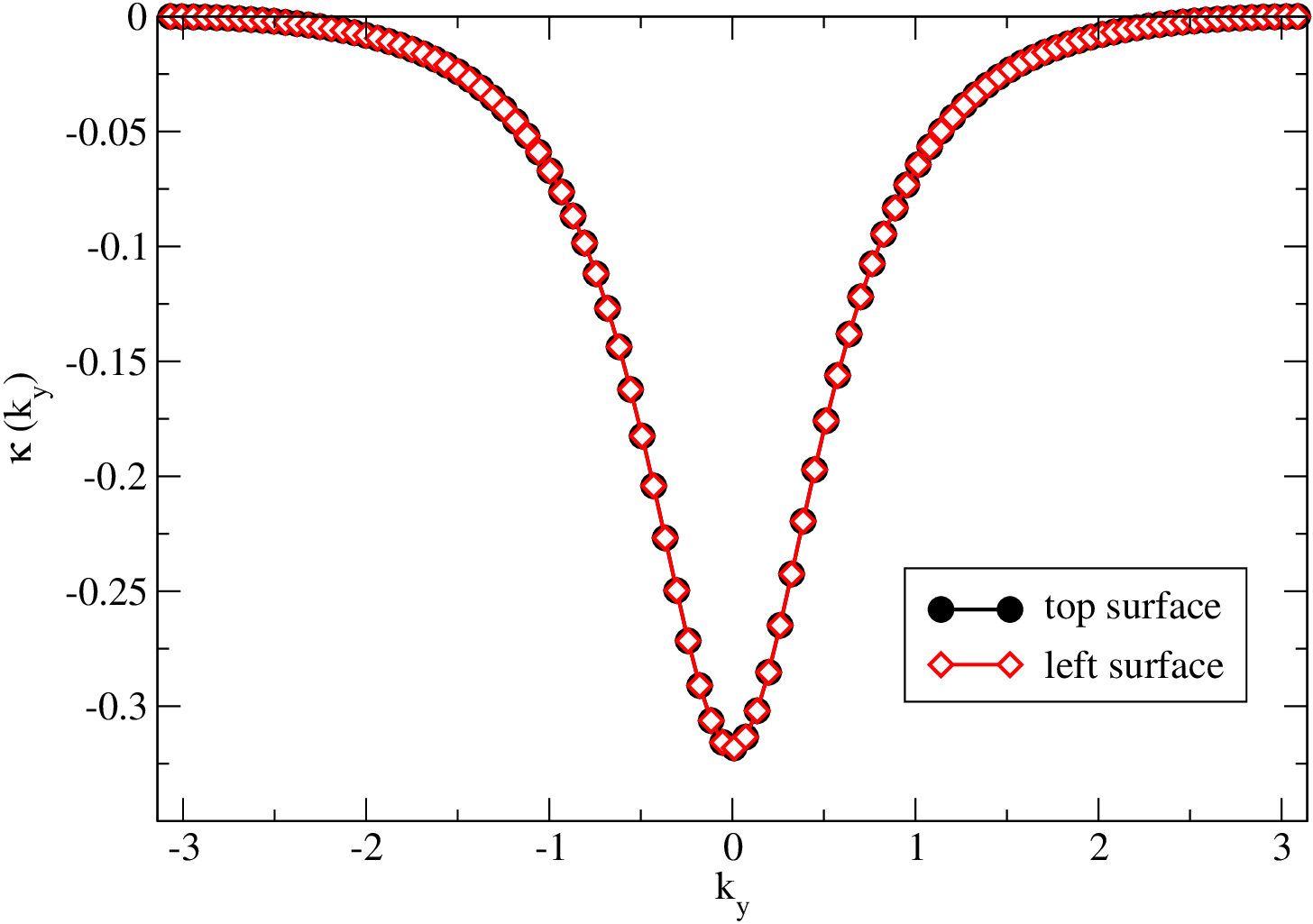}
\caption{(Color online)
$k_y$-resolved current response for the top and side surface.
The current has the same value for different surfaces for each $k_{y}$.
$c=1$, $c_z=1.3$, $m = -0.4$.
}\label{kappa-ky}
\end{center}
\end{figure}

Now, we study the current response on different surfaces.
We shall focus on the $-x$ surface as a side surface
({\it i.e.} surface normal pointing from the sample to its outside along $-\hat x$ ).
The current $I^z_{side}$ generated by $B_{y}$ on this side surface is along the $-z$ direction
and calculated from a formula analogous to Eq. (\ref{Itop})
with $x$ and $z$ interchanged.
In this case, we apply an open boundary condition along the $x$ direction and
periodic boundary conditions in the $y$ and $z$ direction.
Though the calculations in this subsection is done for a film geometry
which is different from those in the last (Sec. \ref{p-v}), our calculations show that
the current response is identical for the same given parameters $c$ and $c_z$
(Fig. \ref{kappa-m}).
This is expected, since for a sample with finite size in both $x$ and $z$ directions,
we must have a (number) current circulating counterclockwise along the $y$ direction.
The currents on different surfaces must be equal, which is inevitable due to the charge conservation.
Even more we find, as shown in Fig. \ref{kappa-ky}, the current is conserved for each $k_{y}$.
($\kappa (k_y)$ is
defined as by Eq. (\ref{kappa}) but with the numerator replaced by
$I^x_{top}(k_y)$ or $I^z_{side}(k_y)$, which are in turn given by a formula similar to
Eq. (\ref{Itop}) except there is no integration over $k_y$.)
It is easy to get this conclusion for $k_{y}=0$, because the system
reduces to a two-dimensional quantum spin-hall system.
For finite $k_y$, one has a tight-binding model for a fictitious two-dimensional
system with time-reversal-symmetry breaking terms.
However, the currents must still be the same for different surfaces again due to the charge conservation.
Our results give a numerical demonstration that current for each momentum
$k_{y}$ is conserved.
This conclusion applies to the cases with surface magnetic moments,
which is discussed further in the next subsection.

Figure \ref{kappa-ky} also shows that the current response is mainly from small $k_{y}$
and decays as increasing $|k_{y}|$.
We note that $\kappa (k_y)$ is a smooth function in $k_y$, even though
surface states exist only for $|k_y|$ below some maximum value (where
the surface states merge into the bulk states).  The sum of the
surface and scattering states contributions give rise to smooth $k_y$ dependence.

The current responses of the side surface for the cases with different $m$ and
anisotropy are also the same as those of the top surface for each given $k_{y}$.
The resulting $\kappa$'s are the same between different surfaces,
as mentioned in the caption of Fig. \ref{kappa-m}.
However, the dependence of the current response on $k_{r}$ is different.
Numerical results for small momenta show that $\kappa$ on the side surface is
weakly enhanced by increasing $c_{z}$ and strongly suppressed by increasing $c$.
This might be explained by the shrink of the momentum region to have the surface states
and the change of slopes of the surface cones. However, this momentum region is
related to bulk properties of a topological insulator by material parameters,
as discussed in the last subsection. In addition, anisotropic effects of
the current response to $B_{z}$ on side surfaces are not
simply explained by the modification of the surface cones.
We have more discussion about this point in Sec. \ref{field-type}.

\subsection{Surface magnetic moment}\label{smm}

Even with the independence of the chemical potential, one may question
if the current changes dramatically when the surface states are gapped out,
for example in the presence of
an extra term in the Hamiltonian
proportional to $s_{z}$, representing local moments, at the top surface.
For clearer explanation, we shall focus on the isotropic case ($c=c_z=1$) and
study the current response to $B_{y}$ on the top surface.
The responses to other fields on different surfaces can be found by a mapping
discussed in the next section.

Before calculation, it is helpful to apply some simple arguments to find out
what we should expect.
Consider surface magnetic moments put just on the top layer. If the size of the system is
large enough in the $z$ direction, the surface current for the bottom
would not be affected. Our numerical study has verified this point for
all considered cases of this paper.
Using charge conservation, the current at the top surface cannot change either.
(This point may be clearer if one considers a sample with finite (but both large) sizes
in both the x and z directions, since, as mentioned in the last section, we
must have a circulating current).
Hence, the current at the top surface cannot be affected by the surface
magnetic moment, even though a gap may appear.  (The same
conclusion can be reached by consider the currents at the side surfaces.)
We shall verify this below by explicit calculations.

\begin{figure}
\begin{center}
\includegraphics*[width=150mm]{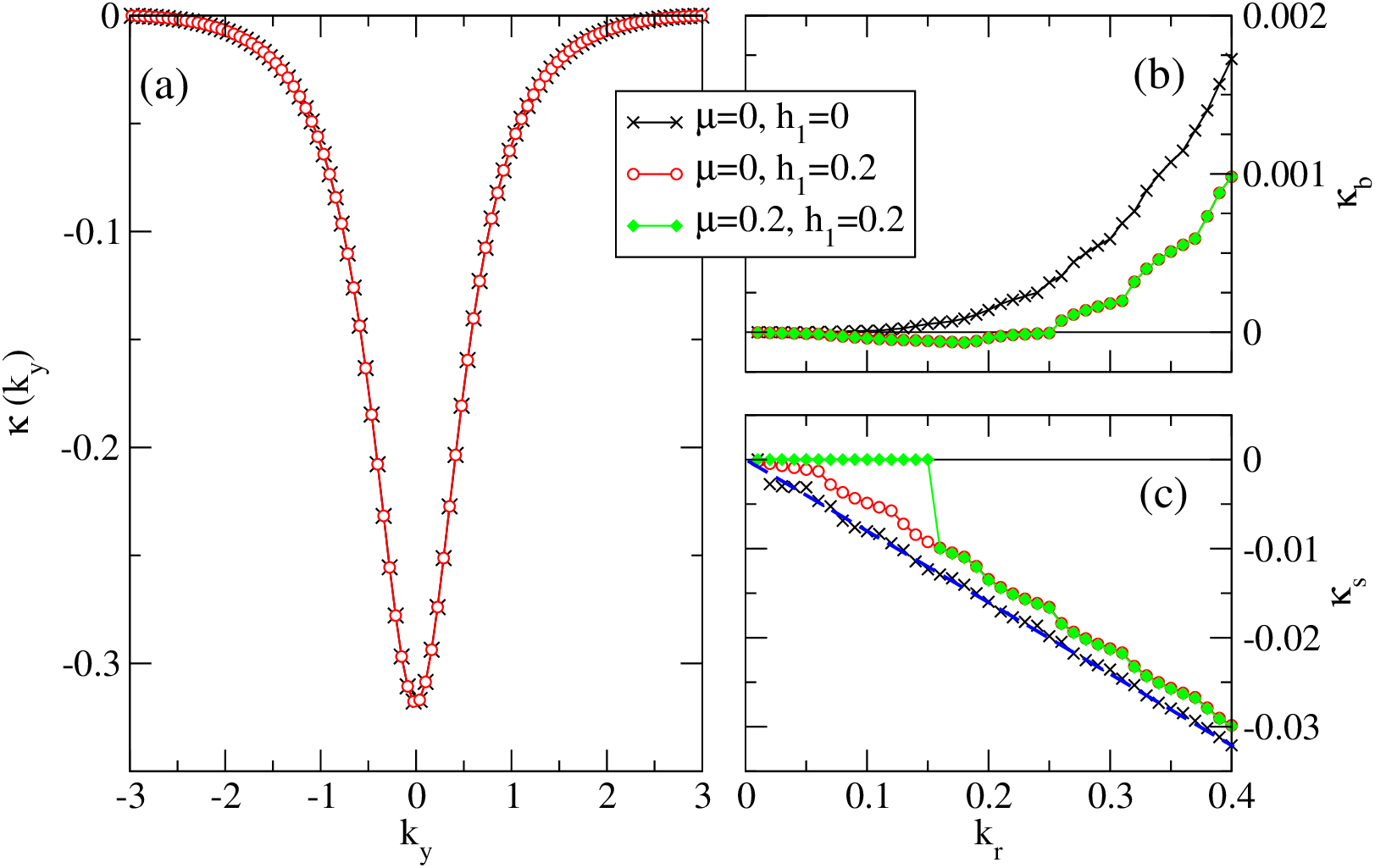}
\caption{(Color online) Current response for the system with surface magnetic moment $h_1$.
(a) The robustness of the current to local magnetic moments for each $k_{y}$.
The contributions from the scattering and the surface states are
shown in (b) and (c).  Same symbols are used for all panels.
The blue dash line in (c) is linear fitting around small $k$ for the case
without surface magnetic moment and with $\mu=0$.
$m= -0.4$.
}\label{kappa-mi}
\end{center}
\end{figure}

To include local magnetic moments at the surface,
we add an additional term
\be
H_{M}=\bf{h_1}(\bf{r})\cdot \bf{s}
\ee
to the Hamiltonian, where $\bf{h_1}(\bf{r})$ is an position dependent exchange field.
(This is not the most general form (see Sec. \ref{field-type} below)
but is sufficient for our considerations here).
We first consider an $\bf{h_1}(\bf{r})$ to produce a gap on the surface states.
For this we use
${\bf h_1}({\bf r}) = h_1({\bf r}) \hat z$ with $h_1$ uniform and finite only
on the top layer of our system.  (The gap generated is
however smaller than $h_{1}$ by a factor from the
overlap between the surface state wavefunction with the first layer).
Numerical results of the current responses for various values of $h_1$
are shown in Fig. \ref{kappa-mi}. Figure \ref{kappa-mi}(a) shows that the total current
response is unchanged:  in fact, the $k_{y}$-resolved current is also independent of $h_1$.
(Note that while Fig. \ref{kappa-ky} presents the current conservation between different surfaces,
Fig. \ref{kappa-mi}(a) shows the robustness of the current to surface magnetic moments.)

The above however does not imply that the current contribution from each in-plane momentum
unaffected by $h_1$.
We show the $k_{r}$ dependence at small momenta in Fig. \ref{kappa-mi}(b-c).
First, let us focus on $\mu =0$.  With finite $h_1$,
$\kappa_s$ is suppressed.  This can be easily understood
by considering the modifications to the surface state spectrum
when a gap is opened up.   On the other hand, $\kappa_b$ also decreases,
and in fact can even change sign at small $k_r$.  These changes compensate
each other exactly when integrated over all momenta, giving a response
independent of $h_1$.
Figure \ref{kappa-mi} also contains a comparison between $\mu =0$ and $\mu = 0.2$ at
a finite $h_1$.  As in Sec. \ref{p-v}, $\kappa_s$ vanishes exactly for
$k_r$ below the Fermi momentum, implying once more that the Van Vleck
contributions involving virtual transitions between the surface and bulk bands
vanish even in the present case, even though the boundary
condition at the surface becomes different from that of Sec. \ref{p-v}
since now there is a finite $h_1$ at the surface.   When $k_r$ is increased beyond the Fermi momentum,
the Pauli contribution again compensates exactly the missing surface-to-surface
Van Vleck contribution, as in Sec. \ref{p-v}.

From the physical argument given in the beginning of this subsection
(or the direct evaluations above),
the independence of the current on the chemical potential in Sec. \ref{p-v}
is then no surprise.  Consider again a given sample with
magnetic moments on the top surface, gapping out the surface states.
The current at the top surface should be independent of $\mu$
if $\mu$ is within the created gap.  By charge conservation,
the same must then be true also for the "pristine" bottom surface.
One can consider the large $h_1$ limit where the surface states disappear
entirely for the top surface and conclude that the total current
at the bottom (and hence for all surfaces) must also be independent of $\mu$.
The same arguments can be applied to the side surfaces when considering
a sample with finite (but large) extent in both $x$ and $z$ directions.
In this case we must have circulating current of equal magnitudes
near both the $x$ and $z$ surfaces when the magnetic field is along the y direction
at all $\mu$'s within the gap and
independent of whether the surface states on one particular surface is destroyed or not.
\footnote{The same argument applies to the current response to an
external orbital magnetic perturbation. Therefore we believe the total current response
(orbital magnetization) is a bulk property.}

\subsection{Non-uniform external fields}\label{non-u}

\begin{figure}
\begin{center}
\includegraphics*[width=150mm]{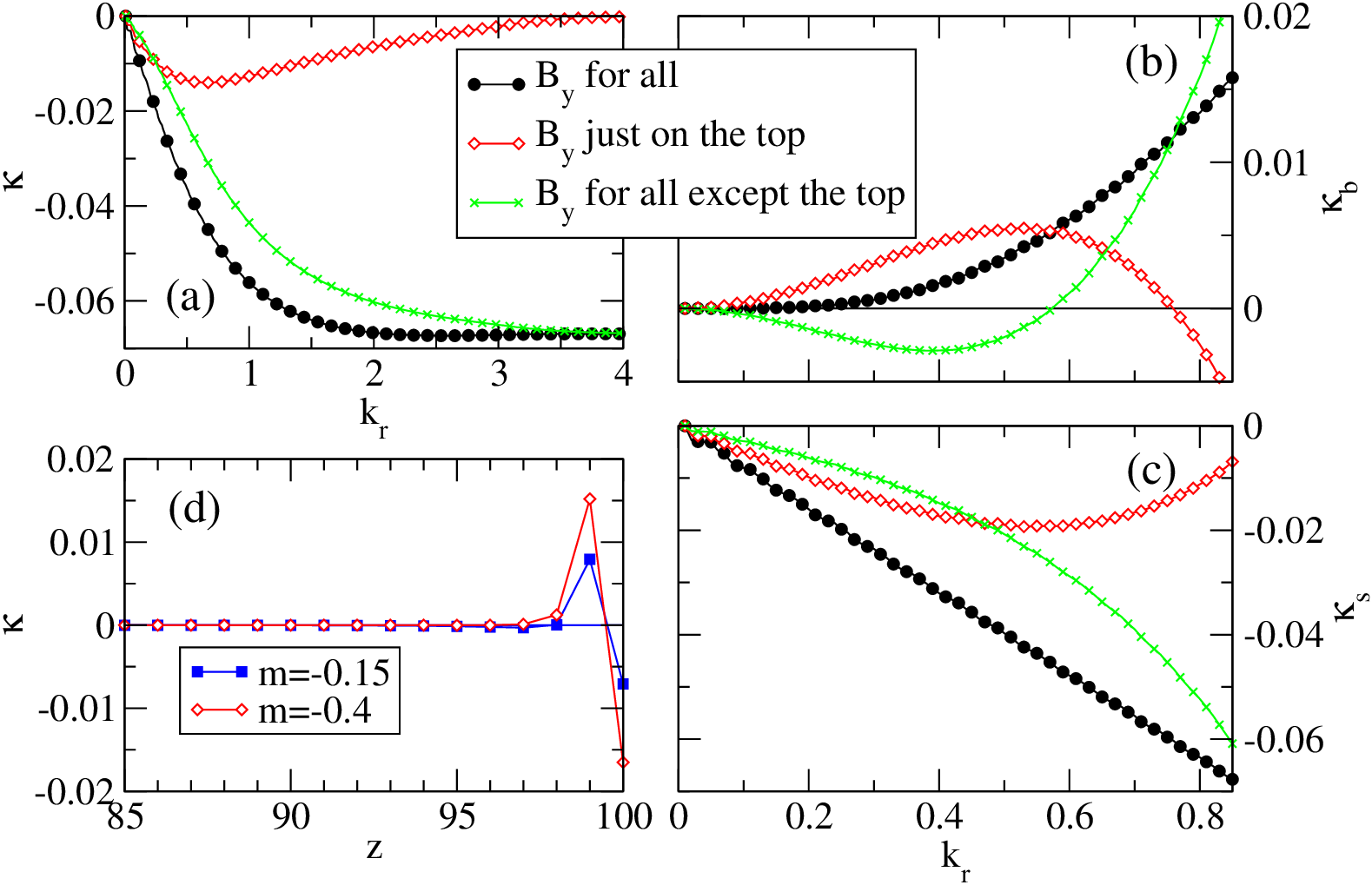}
\caption{(Color online) Current response for the system with position dependent $B_y$.
(a) Total current response for the system with external field for all sites
(black full circle), for all except the top layers (green cross),
and just for the top layers (red empty diamond). The corresponding
contributions from the scattering and the surface states for momenta up to $k_r$
are shown in (b) and (c), respectively. (d) Position dependence of current
response with different mass to $B_{y}$ just on the top layer.   $m= -0.4$ for (a)-(c).
}\label{kappa-mi2}
\end{center}
\end{figure}

This robustness of the current response to a time-reversal-broken term also applies to other
situations.
We shall discuss two special cases, both related to distributed external fields.
The first is an artificial case with an external field on all sites except the top layer.
The results are shown in Fig. \ref{kappa-mi2}(a-c) by lines with crosses.
This case can also be considered a system with surface magnetic moment
with ${\bf h_1}$ finite only on the top layer and along the $-y$ direction
with an magnitude equal to that of the external Zeeman field $B_y$.
Therefore, it is instructive to compare this case with Sec. \ref{smm}.
Comparing Fig. \ref{kappa-mi2}(b-c) with Fig. \ref{kappa-mi}(b-c), we indeed find similar
dependence of the responses on $k_{r}$.
At small $k_r$, $\kappa_s$ is suppressed, but $\kappa_b$ also decreases,
which in fact can also change sign near small $k_r$.
Their sum $\kappa$ at small $k_r$ is smaller, but
the loss at small momenta
is complemented by the contribution from large momenta, which mainly from the scattering states.
After summing over all states, the total current is not affected as shown in Fig. \ref{kappa-mi2}(a).

The second case is a system with $B_{y}$ just on the top surface.
Because one layer is very small as comparing to the size of the bulk,
we can expect no total surface current (since the current at the
bottom surface is necessarily zero).
Indeed as shown in
Fig. \ref{kappa-mi2}(a), we obtain vanished $\kappa$ after summing over
all states.  However, when $k_\|$ is summed only up
to a finite range $k_r$, it is in general finite.
$\kappa_s$ at finite $k_r$ is similar to the uniform $B_y$ case,
being always negative but smaller.  It is somewhat difficult to
separate clearly the surface versus bulk states when $k_r$ becomes
comparable to the $k_\|$ where these states merge into each other,
but our calculation shows that $\kappa_s$ seems to vanish when summing
over all surface states.  We understand this as
follows.  We note that at the parallel momentum where a
surface merges with the bulk states, the decaying length of the surface
state diverges, and therefore a finite localized magnetic moment at the surface
will result only in a vanishing perturbation.  The vanishing
of the current from the surface states can then be understood
as due to the fact that now the surface state spectra simply
shifts in momentum space with their end points fixed.
(Note that this shift in momentum is therefore $\vec k$ dependent and
is different from that from a uniform gauge field).
From Fig. \ref{kappa-mi2}, $\kappa_b$ within $k_r$ is also in general
finite.  In particular, it acquires a negative value near the momentum where
the surface states merge into the bulk, though the total $\kappa$ integrated
over all momenta again vanishes.

Since we are evaluating the linear response, we can see from Fig. \ref{kappa-mi2}
that the response for a uniform $B_y$ is just the sum of
the responses for the two non-uniform $B_y$ cases discussed above,
as it should be.   We have also checked that, for a finite $\mu$, there is
no response for $k_r$ smaller than the Fermi momentum
(as in Fig. \ref{kappa-kr}(c) and \ref{kappa-mi}(c)),
suggesting again that there are no Van-Vleck contributions due
to surface to bulk virtual transitions for our non-uniform $B_y$'s.

Figure \ref{kappa-mi2}(d) shows the position dependence of the total current response
for the case where $B_y$ is localized at the surface.
There is a sharp (lattice-scale) feature near the surface.  This is followed by
a slower feature of opposite sign decaying into the bulk,
with an $m$-dependent decaying length (around $2a$ for $m=-0.4$ and $6a$ for $m=-0.15$.)
This pattern then is roughly a current loop around the localized external field $B_y$.
For a thick sample, there is thus no net current on the bottom surface.

\section{Types of perturbed Hamiltonian for external fields on different surfaces}\label{field-type}

When comparing the current responses to external fields with different directions,
we should take more care about the form of the perturbation Hamiltonian.
Actually, Eq. (\ref{hb}) or Eq. (\ref{hb-lattice}) already implied the
anisotropy of topological insulator between the $z$ and in-plane directions.
For this system, indeed the current response to $B_{y}$ is from $H_{B1,y}$.
If the applied field is along the $z$ direction, however, the current response
of a side surface is not simply from $s_{z}B_{z}$ with some coefficient.
In this section, we shall have more discussion on this point.

We can limit possible terms in the Hamiltonian by symmetry arguments.
Since an external magnetic field is a pseudo-vector
which also breaks the time-reversal symmetry, the allowed terms in
the Hamiltonian describing the effects of a magnetic field can be \cite{Liu10}
\be
H_{B} = -\bf{b}_{1}\cdot\bf{s} - \bf{b}_{2}\cdot\bf{s}\sigma_{x},
\label{h-mi}
\ee
where $\bf{b_{1}}$ and $\bf{b_{2}}$ are external magnetic fields
scaled with appropriate directional dependent $g$-factors.
We shall call the term in Eq. (\ref{h-mi}) without parity operator $H_{B1}$ and the other $H_{B2}$.

\begin{table}[t]
\caption{\label{table} Current response of a topological insulator
and reaction of its surface states to different types of external field on different surfaces.
Sh$_{\pm i}$: surface cone shift in $\pm k_{i}$ direction. Sl: Slope reduced for all momentum.
Sl$_{ex}$: Slope reduced except along $k_{x}=0$.
Sl$_{ey}$: Slope reduced except along $k_{y}=0$.}
\begin{center}
\begin{tabular}{|c|c|c|c|c|}\hline
component       & current response & reaction of        & component       &   reaction of \\
on $+z$ surface & on $+z$ surface & top surface states & on $+x$ surface & side surface states \\ \hline
$s_{x}$ & yes & Sh$_{-y}$ & $s_{y}$ & Sh$_{-z}$ \\ \hline
$s_{y}$ & yes & Sh$_{x}$ & $-s_{z}\sigma_{x}$ & Sh$_{y}$ \\ \hline
$s_{z}$ & no & Gap & $-s_{x}\sigma_{x}$ & Gap \\ \hline
$s_{x}\sigma_{x}$ & no & Sl$_{ey}$ & $s_{y}\sigma_{x}$ & Sl$_{ez}$ \\ \hline
$s_{y}\sigma_{x}$ & no & Sl$_{ex}$ & $-s_{z}$ & Sl$_{ey}$ \\ \hline
$s_{z}\sigma_{x}$ & no & Sl & $-s_{x}$ & Sl \\ \hline
\end{tabular}
\end{center}
\end{table}

Our numerical calculations show that, for the top surface, current
can be generated from $b_{x}$ and $b_{y}$ by $s_{x}$ and $s_{y}$, respectively.
On the other hand, non-zero current response on the x-side surfaces is
from $b_{y}$ and $b_{z}$ by $s_{y}$ and $s_{z}\sigma_{x}$, respectively.
These conclusions agreed with those of \cite{Yip14} which analyzes only states
with small momenta.
The responses might be possibly be understood by considering the reaction of the surface states
to the external fields, which is summarized in Table \ref{table}.
Therefore, it might be reasonable to expect that the system has a
current response as the surface cone is shifted by external fields.
For example, the component $s_{y}$ shifts the surface cone in $+k_{x}$ direction
and creates a contribution of the number current in $-x$ direction if the
perturbed Hamiltonian is $H_{B1,y}$.
Also, $s_{y}$ shifts the surface cone in $-k_{z}$ direction for $+x$ surface.
Therefore $H_{B1,y}$ will create a surface current along $+z$ direction on the x-side
surface, as discussed before.
Those denoted by gap or reduced slope have no current response.
For example, another possible term from $B_{y}$ is $H_{B2,y}$, which excite
no current on both the top and side surfaces.
Even though for the total physical response one must consider
both the surface and bulk states,
it turns out that the above criteria can decide
whether there is or is not a finite response to a particular external perturbation.

\begin{figure}
\begin{center}
\includegraphics*[width=100mm]{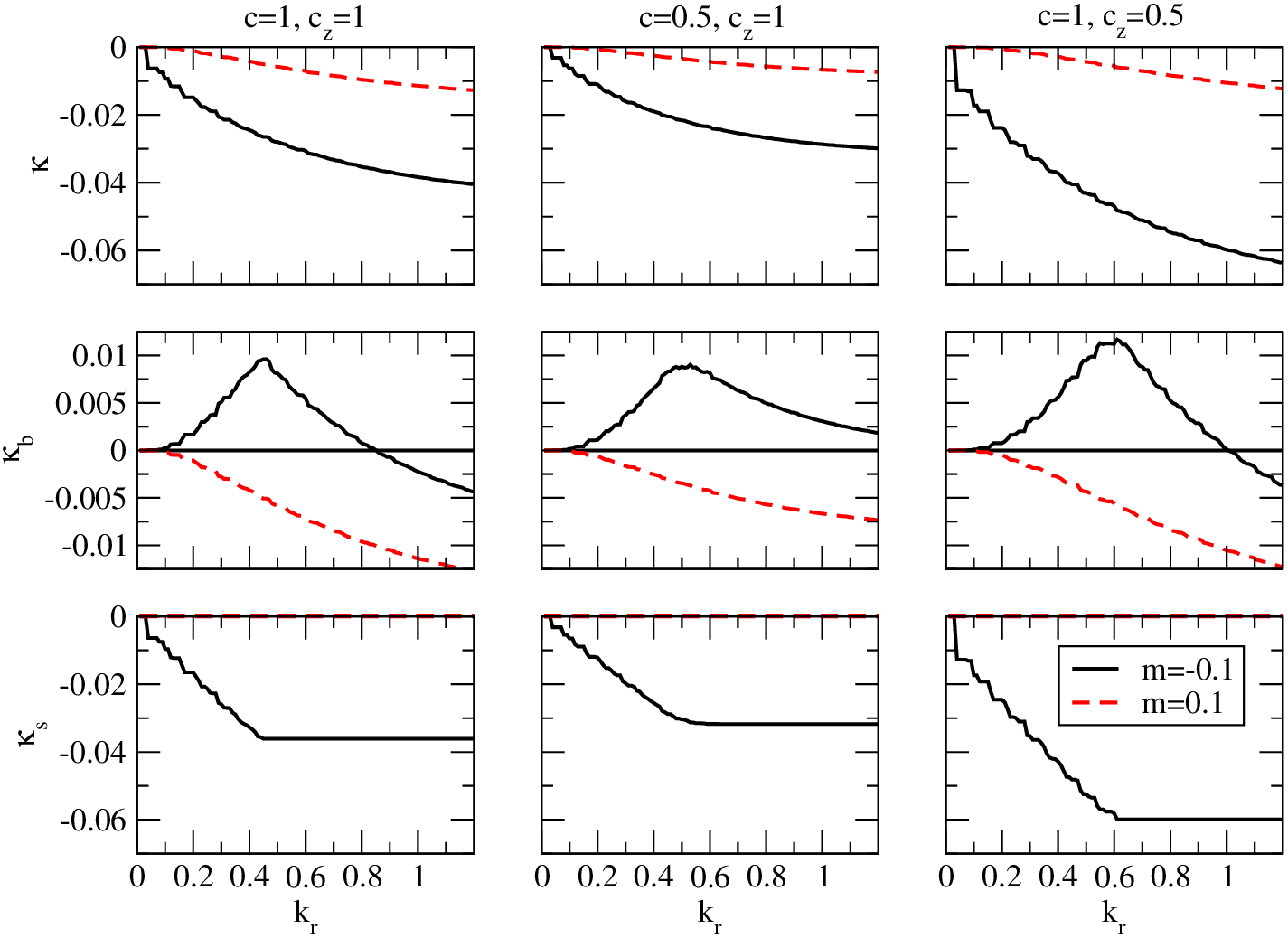}
\caption{(Color online) Anisotropy study of the current responses to $B_{z}$
on a side surface. The arrangement of these plots is the same as
that illustrated in Fig. \ref{kappa-aniso} except
the external field is along the $z$ direction.
}\label{kappa-aniso-side}
\end{center}
\end{figure}

As a comparison, we like to demonstrate anisotropic effects of the current responses
to $B_{z}$ on side surfaces.
The maximum momentum region to have surface states within the constraint
Eq. (\ref{hb-sscondition}) is similar to an ellipse as $c\neq c_{z}$.
The radius of this region in the $k_{z}$ direction is
set to be $\rho_{z}$ and that in the $k_{y}$ direction $\rho_{y}$.
From our numerical calculation, we find that, for systems with fixed $c$,
$\rho_{z}$ is shorter while $c_{z}$ is larger.
For systems with fixed $c_{z}$, increasing $c$ will decrease $\rho_{y}$.
The parameter change helps to understand the current response of systems
with these material parameters.
The first and third columns of Fig. \ref{kappa-aniso-side} show that
$\kappa$ decreases as $c_{z}$ becomes larger.
This might be explained by the shrink of $\rho_{z}$, which supposed to reduce $\kappa_{s}$.
$\kappa_{s}$ no longer contributes to $\kappa$ when $k_{r}$ is larger than some critical value,
but the curve of $\kappa$ smoothly increases as crossing the transition.
$\kappa_{b}$ compensates the loss in the region without those midgap states.
On the other hand, the first two columns of Fig. \ref{kappa-aniso-side} show that
$\kappa$ becomes larger when $c$ is larger.
One might relate this increasing to sharper slope of the surface cone in $k_{y}$.
However, the shrink of $\rho_{y}$ would make the surface states have less region to contribute $\kappa_{s}$.
The same reduction occurs for $\kappa_{b}$.
After summation, $\kappa$ still smoothly increases as including more contribution from larger momenta.
The current responses on a side surface cannot be simply explained from the study of the surface states.
Besides, the momentum region to have surface states is determined by the material parameters,
which are used to describe the bulk properties. It is necessary to consider
both the surface and the scattering states in current responses.

\section{Conclusion}\label{conclusion}

In conclusion, we studied the current response of a topological insulator
to a static Zeeman field using an exact diagonalization to a lattice model.
The effects of current responses to different types of external
fields on different surfaces are discussed.
We find no change in the surface current by changing the occupancy of the
surface states or applying a time-reversal symmetry-breaking terms near the boundary.
These only affect the individual contributions to the current,
such as Pauli versus Van-Vleck, or surface versus bulk states, but
not the total.  This suggests that there is a kind of sum-rule for this response,
though we have not yet been able to derive it analytically.  \footnote{%\cite{footnote-W}
We note here that there are recent reports \cite{2D} of constructing localized
Wannier functions for two-dimensional topological insulators, which would help in understanding
this question. For three-dimensional topological insulators, this construction has not been reported,
though there is a claim \cite{Panati07,Brouder07} that this should be feasible.}
Our results support the conclusion that magnetization is a bulk property,
independent of the details at the boundaries of the sample, even for
topological insulators.

\ack
This work is supported by the Ministry of Science and Education of Taiwan
under grant number MOST- 101-2112-M-001 -021 -MY3.

\end{document}